\documentclass[seceq]{ptptex}

\usepackage{graphicx}
\usepackage{wrapft}

\title{Dynamical Stability of Cusps and Cores: \\
Implications to Centers of Galaxies and Clusters of Galaxies}

\author{
Tadashi \textsc{Nakajima}$^{1}$\footnote{E-mail: tadashi.nakajima@nao.ac.jp}
\footnote{Version \today}
}

\inst{$^{1}$National Astronomical Observatory of Japan, 
Osawa 2-21-1, Mitaka, 181-8588, 
Japan 
}

\abst{
We study the stability 
of a spherically symmetric density profile.
First, we examine the case for ordinary gas 
with a well-defined equation of state (EOS).
In the Newtonian regime, the analysis is based on
the equation of hydrostatic
equilibrium, and in the general relativistic regime, it is based on 
the Tolman-Oppenheimer-Volkoff equation.
We find that a stable system must have
a flat top, or $d\rho/dr = 0$ at $r=0$.
For a power-law equation of state,
$p \propto \rho^\beta$, $\beta > 0$ is required in the Newtonian regime,
while $\beta > 1$ is necessary in the relativistic case.
At least, some contribution from degeneracy  pressure of fermions ($1<\beta \le 5/3$) 
is necessary to support a stable structure in the case of relativistic gas.
In these cases, stability conditions are justified
thermodynamically as well as dynamically. 
Second, we analyze the case of a collisionless system with
a power-law profile given by $\rho \propto r^{-\alpha}$,
in the Newtonian regime using the Jeans equation.
The Jeans equation is mathematically identical to the equation of
hydrostatic equilibrium as long as $p_J = \rho \sigma^2$ is 
regarded as the pressure counterpart for ordinary gas,
where $\sigma$ is the local velocity dispersion.
Since a self-gravitating collisionless system is at best
metastable, if it is treated statistical mechanically,
the dynamical equilibrium described by the Jeans equation will
have a shorter life time than the thermodynamically stable equilibrium
for ordinary gas. Here we take a heuristic approach, using
the pressure counterpart, $p_J$. From the requirement
that $p_J > 0$ or the normalcy of the formal EOS, $p_J = p_J(\rho)$
obtained by eliminating $r$, we find the following heuristic
stability criterion. 
For $0 < \alpha < 1$, $p_J < 0$, the density profile
is unstable. We interpret this profile to be a transient one
evolving toward a flat-top profile, which seems to be the final product
with absolute stability.
For $1<\alpha <3$, the profile is dynamically stable.
It is an interesting coincidence that dark matter halo profiles
obtained by cosmological N-body simulations, fall into this category.
For $3>\alpha$, $p_J<0$. We suspect that such a steep
profile is in the process of collapsing toward $r=0$.
Finally, our stability criteria are compared with
the density profiles obtained for galaxies
and clusters of galaxies.
}

\begin{document}



\maketitle

\section{Introduction}

For ordinary gas, with a well-defined equation of state (EOS),
the equilibrium structure of 
a spherically symmetric self-gravitating system such as a star,
can be obtained by integrating the equation of hydrostatic equilibrium (HSE)
in the Newtonian regime and the Tolman-Oppenheimer-Volkoff (TOV) equation
\cite{OV39}
in the relativistic regime. In this process, both inner and outer boundary
conditions must be specified. 
At the inner boundary, both the pressure, $p$, and density, $\rho$,
need to be positive and finite. By integrating the equation of HSE or
TOV equation, we show that these requirements on $p$ and $\rho$ lead
to the requirements on the flat-top nature, $d\rho/dr=0$ at $r=0$,
of the density profile and on the power-law index of 
the EOS, $\beta>1$, for $p \propto \rho^\beta$.
A physical interpretation of the requirement on $\beta$ is 
that at least some contribution of degeneracy pressure of fermions
as a repulsive force is necessary to support a dynamically stable
structure. In the real universe, neutron stars, white dwarfs, and cool
brown dwarfs, are supported by degeneracy pressure of neutrons
or electrons. 

Now we turn to the analysis of a spherically symmetric system composed of
collisionless particles in the Newtonian gravity \cite{Binney}.
There is mathematically a formal analogy between the equation of HSE
and the isotropic Jeans equation, if we
define the pressure counterpart $p_J = \rho \sigma^2$,
where $\sigma$ is the local velocity dispersion.
The Jeans equation describes the situation in which
the radial gradient in the product of  the density and velocity variance is
balanced with the gravity induced by the encircled mass at a given $r$.
As is well known, statistical mechanics of a self-gravitating
system is not a well-defined subject due to the long-range
nature of the Newtonian gravity \cite{Padmanabhan}.
The equilibrium structure described by the Jeans equation is
at best metastable thermodynamically. On the other hand,
the dynamical equilibrium is realized within a much shorter
time scale than in a thermodynamical relaxation time,
and the Jeans equation is expected to describe the stability
on a time scale much shorter than the thermodynamical stability.

To proceed with the analysis of the stability of 
a density structure described by the Jeans equation,
we make a couple of assumptions regarding the
analogy between the equation of HSE and the Jeans equation.
First, the inner boundary condition imposed on the EOS
for the ordinary gas, is the same as that for the formal
EOS introduced for $p_J$ for a collisionless system.
Second, there is an analogy between 
the thermodynamical normalcy of the gas, which appears
in the behavior of the EOS, and that in the formal EOS 
for the collisionless particles. For instance,
we regard a negative $p_J$ as an indicator of strong
instability of the apparent equilibrium described by the Jeans equation.

We then apply this approach to
a structure described by a power-law density profile
given by $\rho \propto r^{-\alpha}$.
At a finite $r$, elimination of $r$ in the Jeans equation
gives an ``EOS'' or the relation between $p_J$ and $\rho$.
If our criterion turns out to be correct, 
it becomes possible to describe
the strength of ``metastability'', or ``instability'',
which depends on the value of the power-law index,
$\alpha$.

We admit that these assumptions are not fully backed up by
a sound physical principle. 
Our justification for now, is more historical than scientific.
In the past, some heuristic approaches proceeded
to the development based on rigorous physical principles.
What we hope here is that if we find some
sound physical basis of this approach
in the future, our stability criterion will
become more reliable, since
the  results of mass distribution measurements of  
astronomical objects and cosmological simulations,
are often expressed by this power-law form.

The paper is organized as follows. We first elaborate
the derivation of the stability criteria for ordinary gas
in \S2. Our main discussion on the heuristic stability analysis
for collisionless particles is given in \S3. 
The applicability of our stability criterion to centers of galaxies and
clusters of galaxies, and 
density profiles obtained by N-body simulations, is discussed in \S4.

\section{A Necessary Stability Condition for a Self-Gravitating Gaseous System}

\subsection{Non-Relativistic Regime}

The basis equations are the equation of hydrostatic equilibrium (HSE),

\begin{equation}
\frac{dp(r)}{dr} = -\rho(r) \frac{GM(r)}{r^2},
\label{HSE1}
\end{equation}

combined with the encircled mass relation, 

\begin{equation}
dM(r) = 4 \pi r^2 \rho(r) dr.
\label{HSE2}
\end{equation}

In the case of a star without nuclear energy generation, these equations are simultaneously integrated,
starting from either the inner boundary or the outer boundary,
for a given equation of state (EOS),
$p = f(\rho)$.  For a dynamically stable structure,
there is a necessary condition imposed at the inner boundary at $r=0$,
that $p$ is positive and finite at $r=0$.

Now we consider the case that the EOS is given by a power-law as,

\begin{equation}
p = A \rho^\beta.
\label{EOS1}
\end{equation}

As long as $\beta$ is finite, $\rho$, is also finite at $r=0$,
which gives the boundary condition for $M(r)$ as $M(0)=0$.

Eq(\ref{HSE1}) for a small $r$ can be expressed as, 

\begin{equation}
\frac{dp}{dr}  = -4 G \pi \rho(0)^2 r.
\label{HSE3}
\end{equation}

Then, $dp/dr \rightarrow 0$ as $r \rightarrow 0$ and
$dp/dr < 0$ for a finite $r$. From the same argument,
we obtain the similar behavior of $\rho$
for positive $A$ and $\beta$. 

Therefore, the stability near the origin requires
the following two conditions,

\noindent (1)
The density structure must have a flat top.

\noindent (2)
For the EOS, $p$ and $\beta$ must be positive. This means that
$p$ increases as $\rho$ increases, indicating  thermodynamical normalcy.

\subsection{General Relativistic Stability and Equation of State}

At extreme high densities near the origin of symmetry of mass distribution,
we have to take into account at least two physical phenomena outside the scope of 
the classical Newtonian mechanics,
the quantum statistical degeneracy, and the
general relativistic treatment of stability.
Even if dark matter particles are only weakly interacting in the sense of
particle physics in vacuum, indistinguishability of identical particles 
affects statistical dynamics,
when the mean particle spacing becomes smaller than 
the thermal de Broglie wavelength. 

Actually this is not exceptionally remote possibility.
At the core of a massive cluster of galaxies, the mass density
reaches $10^{-24}$ g cm$^{-3}$. If a significant fraction of dark matter is
composed of eV-mass massive neutrinos, they are expected to be degenerate
on the scale of 100 kpc \cite{NM06}.

Although a collisionless Newtonian system composed purely of cold dark matter
may be thermodynamically metastable, the effect of repulsion by
degeneracy pressure of neutrinos may be to make the mixture
of neutrinos and cold dark matter particles into a thermodynamically
normal system.

Here we assume that
the equation of state (EOS) of the dark matter gas will be well defined
in the local inertial frame near the origin. The EOS in terms of the pressure law,
$p = p(\rho)$, governs the behavior of the energy momentum tensor, and thus
the stability. 
Again we consider the EOS expressed as a power law as,

\begin{equation}
p = A \rho^\beta,
\label{powerlaw}
\end{equation}

where $A$ does not depend on $\rho$, but may depend on temperature, $T$, and
particle properties such as the particle mass and statistical weight. 
We examine the range of the power law index, $\beta$, for bosons and fermions,
under different physical conditions.
First we list up the extreme cases and then give the ranges of $\beta$ for bosons and fermions
separately.

For nonrelativistic and non-degenerate gas (Maxwell-Boltzmann ideal gas),

\begin{equation}
\beta =  1.   
\label{Boltzmann}
\end{equation}

For extremely relativistic bosons,

\begin{equation}
\beta = 0,
\label{light}
\end{equation}

since they can be regarded as a black body for which particle numbers are not conserved.

For nonrelativistic partially degenerate bosons, 
most particles are in the ground state and remaining 
thermally excited particles contribute to the pressure, $p$. $p$ depends 
on the number of thermal particles determined by
temperature, $T$, but not by $\rho$ or the total number of
the particles \cite{Landau}.

\begin{equation}
\beta = 0. 
\label{degeneratebosons}
\end{equation}

Including the cases with intermediate physical conditions, the range of $\beta$ for bosons, is

\begin{equation}
\beta = 0 \sim 1.
\label{bosons}
\end{equation}

In the case of fermions, the range of $\beta$ for fully degenerate ideal gas is between
4/3 (extreme relativistic) and 5/3 (nonrelativistic) \cite{Landau}.
Including the cases with intermediate degeneracy, the range of $\beta$ is 

\begin{equation}
\beta = 1 \sim 5/3.
\label{fermions}
\end{equation}

This problem setting under general relativity may appear formidable. However it has been dealt with
by the classical paper by Oppenheimer and Volkoff (OV39) on the neutron stellar core \cite{OV39}. They first formulated
the problem based on Tolman's coordinate system for spherical symmetry \cite{Tolman} for a general
equation of state and then applied the particular parametric equation of state for fully degenerate fermions
covering nonrelativistic to extreme relativistic particles \cite{Chandra,Landau} to solve
the neutron star problem, assuming neutrons as ideal fermi gas. 
The general relativistic counterpart of the equation of the hydrostatic equilibrium is
the TOV equation after Tolman, Oppenheimer and Volkoff.
For the generalized radial coordinate
$r$ in the Tolman's coordinate system and corresponding pressure gradient $dp/dr$, 
OV39 discuss the local stability requirement in the footnote. The result is that
the local stability is satisfied for $\beta > 1$. 

Here we briefly summarize and elaborate the argument by OV39.
Using the Tolman's coordinate system and the 
dimensionless encircled gravitational mass (energy), $u(r)$,
Einstein's equations are reduced to two equations,

\begin{equation}
\frac{du}{dr} = 4 \pi \epsilon(p) r^2,
\label{g-energy}
\end{equation}

and

\begin{equation}
\frac{dp}{dr} = -\frac{p + \epsilon(p)}{r(r-2u)}[4 \pi p r^3 + u],
\label{TOV}
\end{equation}

where $\epsilon(p)$ is the energy density and $p$ is the pressure in the proper coordinate of the matter.
OV39 require that at $r=0$, $u$ and $p$ must have $u_0=0$ and a positive finite value, $p_0$ 
respectively. 
If $\epsilon(p)$ is given as a power law of $p$ as, 

\begin{equation}
\epsilon(p) = C_\epsilon p^s,
\label{EOSE}
\end{equation}

near the origin, eq(\ref{TOV}) reduces to 

\begin{equation}
\frac{dp}{dr} = \frac{p + C_\epsilon p^s}{2r}.
\label{TOV0}
\end{equation}

For $p_0$ to be positive and finite, integration of eq(\ref{TOV0}) shows that
$s < 1$.

Now we examine the relation between the power-law EOS, $p = A \rho^\beta$,
and $\epsilon$.

\begin{equation}
\epsilon = \rho c^2 + \epsilon_K,
\label{epsilon}
\end{equation}

where $\epsilon_K$ is the kinetic energy density.

In the nonrelativistic case, $\rho c^2$ dominates $\epsilon$ and 

\begin{equation}
\epsilon \approx \left(\frac{p}{A}\right)^{1/\beta}.
\label{EOSE2}
\end{equation}

So one can see that $s<1$ is equivalent of $\beta > 1$.

In the ultrarelativistic limit in which gas is treated as radiation,

\begin{equation}
p = \frac{1}{3} \epsilon,
\label{rela}
\end{equation}

or $s=1$. Therefore there is no stable solution for ultrarelativistic case.
One may worry that the Chandrasekhar limit corresponds to the EOS of ultrarelativistic 
degenerate electrons in white dwarfs. In the case of white dwarfs, the gravity source is hadrons,
which are nonrelativistic and gravity is Newtonian. On the other hand,
the TOV equation is dealing with ``self-gravity'' of the matter. 
Even at the center of a neutron star, neutrons are only moderately relativistic.

We need to examine more carefully for the moderately relativistic case.
In the nonrelativistic limit,
for ideal gas  regardless of
bosons, fermions or degree of degeneracy,

\begin{equation}
p = \frac{2}{3} \epsilon_K. 
\label{nonrela}
\end{equation}

and from the relativistic limit, eq(\ref{rela}),

\begin{equation}
p =  \frac{1}{3} \epsilon_K \le \frac{1}{3} \epsilon.
\label{rela2}
\end{equation}

So it is apparent that the second term of eq(\ref{epsilon}), $\epsilon_K$ alone
never prevents the structure from collapsing at the origin.
So the stability at the origin is fully determined by the behavior of
$\rho$ as a function of $p$. Therefore $\beta > 1$ is always required.
For $\beta>1$,
$d\rho/dr=0$ at $r=0$ and the integration of the TOV equations gives a flat-top density profile again
at a small $r$. To summarize, the stability condition near the origin requires,

\noindent
(1) The density structure must have a flat top.

\noindent
(2) For the power-law EOS, $p>0$ and $\beta>1$ are required. Since the system is
unstable for $\beta = 1$, classical thermal pressure can never support a relativistic structure.
To form a stable relativistic structure, some contribution from degeneracy pressure by fermions,
or the repulsive force due to Pauli's exclusion principle, is required.

\section{Stability of a Power-Law Density Profile for Collisionless Particles in the Newtonian Regime}

\subsection{A Heuristic Treatment of the Jeans Equation}

Now we turn to our heuristic analysis of a spherically symmetric collisionless system in the Newtonian regime.
In the following, we use the Jeans equation, trying to find some analogy with
the equation of hydrostatic equilibrium for ordinary gas. We formally introduce the pressure counterpart
and the counterpart for the EOS. As is well known, a self-gravitating system formally treated
by the micro-canonical ensemble of statistical mechanics, is thermodynamically metastable, when
the contributions of the kinetic energy and gravitational potential are comparable \cite{Lynden-Bell68,Padmanabhan}.
This is exactly the situation that the Jeans equation is dealing with. So first, we accept
this thermodynamical anomaly, but expect some type of dynamical stability, when the Jeans equation is valid.
Below we show that depending the steepness of the power law, the pressure counterpart becomes negative and
the formal EOS indicates strong anomaly. In this case, we question the applicability of
the Jean equation itself and the corresponding density profile is regarded as ``strongly metastable'' or 
even ``dynamically unstable'', compared to a system with normal EOS under the Jeans equation.
We also caution that our ``pressure'' or EOS approach is not same as the treatment of the statistical mechanics
either. To define a pressure, in the sense of standard statistical mechanics, we must define a boundary
to define a volume. However, this procedure is inconsistent with the long-range and unshielded nature of the gravity,
and therefore ``pressure'' of this sense was never defined by the approach of statistical mechanics.

\subsection{A Power-Law Density Profile}

The results of mass distribution measurements of astronomical objects, such as 
those of galaxies and clusters of galaxies, are often expressed as
density laws in the form of 

\begin{equation}
\rho(r) \propto r^{-\alpha}.
\label{PL1}
\end{equation}

In the following, we examine how much information is extracted from this
phenomenological profile. We  begin with some certain facts, and then
proceed to our heuristic analysis of dynamical stability.

\subsection{Encircled Mass and Velocity Dispersion}

The most simple physical quantity for a given density law, eq(\ref{PL1}), is the encircled mass, $M(r)$.

\begin{equation}
M(r) = \int_0^r 4 \pi \rho(r) r^2 dr \propto r^{3-\alpha}.
\label{EM}  
\end{equation}

So $M(r) \rightarrow 0$ for $r \rightarrow 0 $, only if $\alpha <3$.
However $M(0) = 0$  is only a necessary condition for a real astronomical
object, since even a black hole has a finite extent.
If one finds a profile with $\alpha > 3$ in a certain range of $r$, it indicates the presence of an
extra-mass source inside, indescribable with the power-law alone.

The second quantity to consider is the velocity dispersion, $\sigma(r)$,
which is often an observable. 

\begin{equation}
\sigma \propto \sqrt{\frac{M(r)}{r}} \propto r^{1-\alpha/2}.
\label{sigma}
\end{equation}

Then, $\sigma(r) \rightarrow \infty $ as $r \rightarrow 0 $ for $\alpha > 2$.
In reality, the relativistic requirement, $\sigma < c$, sets the limit on the minimum value of $r_{min}$,
which gives an order estimation of the Schwarzschild radius, $2GM/c^2$.
So if a cusp shows a rise of density towards the center with $\alpha > 2$,
this is an indication of the presence of a compact source inside. 
It may appear that there is no anomaly for $\alpha \le 2$. However,
$\sigma$ simply indicates the behavior of test particles
and it is not necessarily related to the stability.

\subsection{Jeans Equation}

For an isolated spherically symmetric system with collisionless particles, such as that of  cold dark matter
particles or an idealized stellar system with equal mass stars,
the dynamical stability condition is  expressed by the isotropic 
Jeans equation \cite{Binney} as,

\begin{equation} 
  \frac{d[n(r)\sigma^2(r)]}{dr} = - \frac{[n(r) GM(r)]}{r^2}, 
  \label{Jeans}
\end{equation}

with 

\begin{equation}
  M(r) = \int_0^r 4\pi m n(r) r^2 dr,
  \label{Mr}
\end{equation} 

where $n(r)$ is the particle number density, $m$ is 
the particle mass.
  
Suppose we set
 
\begin{equation} 
\rho(r)  =  m n(r)
\end{equation}

and

\begin{equation}
p_J = m n(r) \sigma(r)^2, 
\label{equivalence}
\end{equation} 

where we note that $\sigma(r)$ is the radial component of 
the total velocity dispersion, $\sigma_{tot} = \sqrt{3} \sigma$,
then the Jeans equation and the encircled mass can be rewritten as 

\begin{equation} 
  \frac{dp_J}{dr} = - \rho(r) \frac{GM(r)}{r^2}, 
\end{equation}

and

\begin{equation}
         M(r) = \int_0^r 4\pi \rho(r) r^2 dr.
\label{JEANS}
\end{equation} 

Here we caution that an observable velocity dispersion is often the component 
in the line of sight, $\sigma$, instead of  
 $\sigma_{tot}$.

Therefore,
the Jeans equation for collisionless particles and 
the equation of hydrostatic equilibrium are identical, as long as
we regard 

\begin{equation}
p_J = m n(r) \sigma(r)^2 = \rho(r) \sigma(r)^2,
\label{pJ}
\end{equation}

of a collisionless system
as a counterpart for the pressure, $p$ of ordinary gas. 
This analogy is at least energetically reasonable, since the pressure 
is two thirds of the kinetic energy density for nonrelativistic ideal gas.
However from the point of view of particle momenta or force,
there is a clear distinction. Gas pressure is the total momenta given to
a boundary wall per unit time, while no boundary can be defined for 
the self-gravitating system. So we should regard this analogy
as mathematical for now.
Now we examine the behavior of the power-law profile by the Jeans equation and $p_J$ in detail.

If we integrate eq(\ref{Jeans}), for the power-law profile, eq(\ref{PL1}),

\begin{equation}
p_J(r)  =  A_p r^{2(1-\alpha)} + p_J(0)
\end{equation}

and

\begin{equation}
A_p  =  - \frac{4 \pi G} {(2 - 2 \alpha) (3 - \alpha)}, 
\label{pressure}
\end{equation}

where $A_p$ is a constant coefficient and $p(0)$ is a constant of integration.
When $\rho(r)$ is given as eq(\ref{PL1}), $\sigma(r)$ is also another power law as
eq(\ref{sigma}), then $p_J$ must also be a power law. Therefore $p_J(0) = 0$.
So 

\begin{equation}
p_J(r)  =  A_p r^{2(1-\alpha)},
\label{pressure2}
\end{equation}

where $r^{2(1-\alpha)}>0$ for a finite $r$, and the sign of $p_J$ is determined
by $A_p$ as,

\begin{eqnarray}
A_p & < & 0  \hskip 1cm  {\rm for} ~ \alpha > 3, \\
    & > & 0 \hskip 1cm {\rm for} ~ 1 < \alpha < 3, \\
    & < & 0  \hskip 1cm {\rm for} ~ \alpha < 1. 
\label{Ap}
\end{eqnarray}

Both $\rho(r)$ and $p_J(r)$ are power-laws of $r$ and by eliminating $r$, we
obtain a formal EOS as,

\begin{equation}
p_J = A_p \rho^\gamma,
\label{JEOS}
\end{equation} 

where,

\begin{equation}
\gamma = 2\left(1-\frac{1}{\alpha}\right).
\end{equation}

Now we examine the usage of the pressure counterpart $p_J$ 
and the EOS counterpart, eq(\ref{JEOS}).

\subsubsection{Thermodynamical Instability at the Origin}

From the result of \S2, thermodynamical stability at $r=0$ 
requires, $p_J$ to be positive finite and the density profile
to be a flat top. The former is satisfied only for $\alpha=1$,
but the density profile, $\rho(r) \propto r^{-1}$, does not
have a flat top. So there is no power-law profile
that satisfies ``thermodynamical stability''.
If a flat-top at the origin is combined with a smoothly dying 
density profile near the outer boundary, such as that for a star,
there must be an inflection point of the density profile, somewhere at an intermediate $r$.
On the other hand, a single-power-law profile can never have
an inflection point and the situation is understandable.

\subsubsection{The behavior of the Pressure and EOS Counterparts at a Finite $r$}

At a finite $r$,
if we require the positivity of $p_J$, $\alpha$ must be in the range, 

\begin{equation}
1<\alpha<3.
\label{positivePJ}
\end{equation}

If we further require that the EOS is normal, in other words, $p_J$ increases as
$\rho$ increases, $\gamma > 0$. 
It turns out that this requirement also gives, $1 < \alpha <3$, equivalent to
the positivity condition for $p_J$. 
Since a system described by the Jeans equation is
at most metastable or in dynamical equilibrium with a finite
life time, we consider that $p_J < 0$, indicates even shorter
life time of this stage. 

\subsubsection{Some Conjecture About Physics of Pressure Counterpart}

For $\alpha > 3$, from eq(\ref{EM}),
$M(r) \rightarrow \infty$ as $r \rightarrow 0$. This is an unphysical situation
and a single power-law profile cannot describe a stable structure.
In that sense, $p_J < 0$, may indicate some collapsing stage to the origin.
Or $p_J < 0$ can be an indication that the validity of the static
Jeans equation itself is violated, indicative of dynamical instability.

The range of $\alpha = 1 \sim 3$, for positive $p_J$, has an 
interesting coincidence with the halo density profiles obtained
by cosmological N-body simulations \cite{NFW96,NFW97,M99,F04}.
For instance, the original NFW profile behaves as $\alpha \rightarrow 1$ at small $r$,
while $\alpha \rightarrow 3$, as $r \rightarrow \infty$.
From the encircled mass argument based on eq(\ref{EM}), a finite total
mass at $r \rightarrow \infty$ requires $\alpha > 3$ at a large $r$, and the behavior at a large $r$,
does not seem to carry much dynamical information.
$\alpha \sim 3$, also indicates that the outer region of the NFW profile corresponds to $p_J \sim 0$,
indicative of invalidity of the Jeans equation. 
So our main interest is in the stability of cusps at a small $r$, obtained by N-body simulations.
Including other simulations, the power-law indices of inner cusps are in the range
of $\alpha = 1 \sim 1.5$. This range of $\alpha$ should show a dynamical equilibrium with a long life.

The most interesting region of $\alpha$ is $\alpha = 0 \sim 1$.
As we have seen in \S2, a thermodynamically stable profile must have a flat-top.
The density profile of a star, for, instance, has a flat core,
indescribable with a single power law. If the flat core profile
is the thermodynamically stable final stage, it appears reasonable
to regard this range of $\alpha$ to indicate the transition from
the dynamically stable cusp with $\alpha = 1 \sim 1.5$ to
a thermodynamically stable flat-top core.

The coarse grained view adopted by N-body simulations will give
a proper time scale in the free-fall stage, before dynamical friction sets in.
However once dynamical friction becomes significant at a high density region,
the time scale for dynamical evolution behaves as 
$ N / \log N$ \cite{Padmanabhan}, where $N$ is the total number of collisionless particles
and the particles number or particle mass must be known to 
evaluate the evolution time scale. So the actual duration of
this possible transient stage may not be estimated until we know
the mass of a dark matter particle, even if our conjecture is correct.
The summary of our stability arguments is given in Table \ref{stability-tbl}.

\begin{table}
\caption{Thermodynamical and Dynamical Stability of Density Profiles Based on Our
Criteria\label{stability-tbl}}
\begin{center}
\begin{tabular}{cccccc}
\hline\hline
Profile Type  & $\alpha$  & Thermodynamically & Dynamically &  $p_J$ & comment \\
\hline
Flat Top      &  NA       & Stable          & Stable    &   Positive & Final Stage?  \\
\hline
Power Law     &  $0 < \alpha < 1$ & Unstable      & Unstable  &   Negative & Transient? \\
              &  $1 < \alpha < 3$ & Unstable      & Stable    &   Positive & NFW in this range   \\
              &  $ \alpha > 3 $   & Unstable      & Unstable  &   Negative & Collapsing to origin? \\
\hline
\end{tabular}
\end{center}
\end{table}

\section{Current Observational Status of the Core/Cusp Problem}

\subsection{Limitations in Observations and Applicability of Our Stability Criteria}

Before comparing the observational status with our heuristic stability criteria,
we must carefully examine the limitation in observations to avoid over interpretation.

\subsubsection{Indistinguishability between a Flat-Top Profile and an Unstable Shallow Cusp}

The spatial resolution of observations is always limited and even if 
the true density profile is not a single power law, the observational
results are often fit by a single power law for a limited range of $r$.
For instance, even if the true profile is a flat-top type with
an inflection point, one can fit its observed profile given for
a finite range of $r$  with observational errors, using a single power-law,
and possibly find a shallow cusp with $0 < \alpha <1$.
The real observational situation seems to be that we are not able
to distinguish a flat-top profile from a shallow cusp for external galaxies or
clusters of galaxies.

\subsubsection{Dark Matter, Baryons and Stars}

The density profile of a spiral galaxy is usually obtained by
the measurements of a rotation curve of the interstellar medium.
What rotation curve gives is the encircled mass inside the given radius.
On the average over the entire universe, the mass ratio between
the dark matter and baryons is about 5:1, but
we do not know of this true ratio, especially at the inner region of a
galaxy. The actual presence of stars is a manifestation that
the dissipative collapse and star formation actually happened
in the past. Therefore there is no guarantee that the dark matter
and baryons or stars follow the same density profile.

One possible relief to this complicated situation is that
due to the identity of the mathematical form between the
equation of hydrostatic equilibrium and the Jeans equation,
a similar stability argument may be possible even for the
mixture of ordinary gas and collisionless particles, as
long as we know the total density profile, measured by
the rotation curve. However this is based on a very
optimistic assumption that the dynamical time scales for
dark matter, baryons, and stars, are the same, which is 
not necessarily true. Unless the entire system is  already in dynamical equilibrium,
this  treatment will be invalid.  It is probably wrong, if we analyze the
stability of the system with a mixture of various types of particles,
using our criteria.

This situation seems severe for individual galaxies, while
for a cluster of galaxies, it appears safer to
interpret its overall density profile as representing its dark
matter distribution. This is probably because,
the total stellar mass to the entire cluster mass is
at most a few percent. So a galaxy in the cluster can raise this ratio locally,
but the overall density profile is not significantly affected.

\subsection{Observational Status}

Enormous observational efforts have been made to obtain mass distributions
of the central regions of galaxies and clusters of galaxies.
Below we comment on the observational status from the point of 
view of our argument on the dynamical stability.
In many cases, the inner density power-law indices are poorly 
constrained due to finite spatial resolutions.
As a practical division, we tentatively define a shallow-cusp/core profile 
for $\alpha <1$ as ``a core'' and a stable-cusp profile for $1<\alpha$ as
``a cusp''.
This discrimination is based on our stability criterion discussed previously.
We emphasize our previous caution that a flat-top stable core
is often indistinguishable from a shallow unstable cusp.

\subsection{Centers of Galaxies}

The observational status is better off for spiral galaxies
for which the abundance of interstellar medium allows measurements
of rotation curves using H$\alpha$,
CO, and HI 21cm emission lines. 
However, the limitation in spatial resolution often hampers 
conclusive division between a core and a cusp.

For dwarf galaxies and low-surface-brightness galaxies (LSBs),
the predominance of constant density cores have been claimed
by many authors \cite{Swaters03,deBlok02,Gentile04,Gentile05,JDSimon05},
with varied confidence levels in their conclusions. 
For instance, Swaters {\it et al.}  do not rule out cusp interpretations
based on their analysis of data quality \cite{Swaters03}, while Gentile {\it et al.} strongly
claim the presence of a constant density core for one dwarf galaxy \cite{Gentile05}.
Nuclear activities or the presence of central black holes are generally not reported for dwarf galaxies and LSBs.
If ``a constant density core'' implies ``a flap-top profile'', such a core is probably
thermodynamically stable and a final product of dynamical evolution.

Our Galaxy is a spiral galaxy 
and its rotation curve is well known \cite{Sofue01}.
At $0.1 < r < 20$ kpc,
the rotation curve is nearly flat, indicating an isothermal
density profile of $\rho(r) \sim r^{-2}$.
If our Galaxy were composed purely of collisionless dark matter particles,
this isothermal profile could describe the entire Galaxy, since $\alpha=2$
implies dynamical stability. 
However in reality, the stellar mass fraction appears high in the inner galaxy,
and dissipative collapse of  baryons apparently happened in the past.
The presence of a bulge and a black hole of
$3\times10^7 M_\odot$ at the Galactic center, indicates,
star formation and subsequent black hole formation, near the center.
$\alpha = 2$ is an isothermal profile and this may indicate
the past relaxation of the mixture of baryons and dark matter
on the scale of the entire Galaxy. 

The rotation curve should either rise (the presence of a black hole)
or decline linearly (the presence of a constant core) towards the center,
if it is measured with sufficient spatial resolution.
Clearly both types exist for spiral galaxies in general \cite{Sofue99}.

For elliptical galaxies with little interstellar gas component,
rotation curves are not available. However at least to our knowledge,
the density profile has been obtained
by the measurement of velocity dispersion of intracluster stars in the halo of a cD galaxy,
A 2199 \cite{Kelson02}. They prefer a finite density core than a cusp for the dark halo density
profile.
For E/S0 galaxies which happen to be the lenses of strongly lensed quasars, Treu and Koopmans
have found that the total mass slope is about $\alpha = 1.75 \pm 0.1$,
while the inner dark matter slope is $\alpha = 1.35$, if the stellar velocity ellipsoid is isotropic,
but $\alpha < 0.6$, if the galaxies are radially anisotropic \cite{Treu04}.
Observationally many ellipticals are expected to harbor black holes as dead quasars.

In any case, the application of our stability criteria is limited for galaxies 
for the reasons mentioned above.

\subsection{Centers of Clusters of Galaxies}

Gravitational lensing gives direct measurements of
two-dimensional mass distributions of clusters of 
galaxies and thus is the most promising approach
for obtaining volume density profiles under the
assumption of spherical symmetry.
Especially, strong lensing provides robust 
information on the two-dimensional encircled masses
within tangential and radial critical radii,
with minimal assumptions. Detailed strong lensing results 
obtained for a couple of nearby clusters, CL 0024+1654 \cite{Tyson98}
and A1689 \cite{TB05a},
indicate the presence of soft cores or at most shallow cusps,
although there is a counter example of a steep dark matter profile in
A383, which appears to be formed under the influence of its cD galaxy \cite{Smith01}.
For A1689, Broadhurst {\it et al.} obtained the column density profile for the entire cluster
by combining strong and weak lensing results \cite{TB05b}. 
Their best-fit core-power-law profile
indicates the presence of a finite density core. 
Sand, Treu and Ellis  combined strong lensing and spectroscopy of gravitational arcs
and the brightest cluster galaxy in the cluster, MS 2137-23, and analyzed
the consistency of observations and the predicted cusps by N-body simulations \cite{Sand02}.
They concluded that $\alpha < 0.9$ with 99\% confidence level and preferred a flat profile.
The same group later obtained a mean power-law index $\alpha = 0.52$,
for three clusters with both radial and tangential arcs \cite{Sand04}.
They suggest that the relationship between dark and visible matter
in the cores of clusters is very complex. 
However, except for the case that the distribution of dark matter is
heavily perturbed by that of  luminous matter, 
it appears safe to consider that strong lensing results favor 
soft cores at the centers of clusters.

The observational situations are summarized in Table \ref{tbl-2}.

\begin{table}
\caption{Density profile observations.
Here we define a core by a soft density profile
with $\alpha < 1$, and a cusp by a steeper profile 
with $\alpha > 1$. \label{tbl-2}}
\begin{center}
\begin{tabular}{cccc}
\hline\hline
Object Type  & Core/Cusp & Method &  Comment \\
\hline
Dwarf Galaxies & Cores & Rotation Curve & No BHs \\
Low Surface Brightness Galaxies & Cores & Rotation Curve & No BHs \\
Our Galaxy (Spiral) & Cusp & Rotation Curve & BH \\
Spiral Galaxies       & Cusps/Cores & Rotation Curve & With/Without BHs \\        
E/S0 Galaxies        & Cusps & Lensing & BHs as Dead QSOs \\ 
Cluster Centers & Cores & Lensing & No BH?  \\
\hline
\end{tabular}
\end{center}
\end{table}

\section{Concluding Remark}

In this paper, we have examined the requirements
for the stability of spherically symmetric density profiles,
first for ordinary gas with well-defined EOS,
and then for collisionless particles, with a heuristic approach.
A flat-top profile is thermodynamically stable near the origin,
and appears to be a form of final product of evolution.
In the case of a power-law density profile for a collision system, $\rho(r) \propto r^{-\alpha}$,
we conjecture that dynamical stability depends on $\alpha$.
The system is unstable and possibly in a transient stage toward a flat-top profile,
for $0 < \alpha < 1$.  The system is thermodynamically metastable, but dynamically stable
with a long life for $1 < \alpha < 3$.  It is unstable for $\alpha > 3$, indicating a collapsing
stage to
the center.

One physical way to support a flat-top density structure 
in a cluster of galaxies, is
to introduce degeneracy pressure
of fermionic dark matter. In a companion paper \cite{NM06},
we model the density profile of the center of
the cluster of galaxies, A1689, under the assumption that
some fraction of dark matter is eV-mass
degenerate fermions/neutrinos.

\section*{Acknowledgemens}

The author thanks Masahiro Morikawa for discussion and
critical comments on the eariler draft of this paper.


\begin{thebibliography}{99}

\bibitem{Binney} J. Binney and S. Tremaine, {\it Galactic Dynamics} (Princeton U. Press, Princeton, 1987).


\bibitem{NM06} 
T. Nakajima and M. Morikawa,  accepted for publication in  Astrophys. J. (2006).

\bibitem{Lynden-Bell68} D. Lynden-Bell and R. Wood, Mon. Not. R. Astron. Soc. {\bf 138}, (1968), 495.

\bibitem{Landau}  L.D. Landau, L. D. and 
E.M. Lifshitz, 
{\it Statistical Physics}, (3rd Ed.1980).

\bibitem{OV39}  J.R. Oppenheimer and
G.M. Volkoff, G. M. Phys. Rev. {\bf 55}, 374 (1939), 374.

\bibitem{Padmanabhan} T. Padmanabhan. Phys. Rep. {\bf 188}, (1990), 285. 

\bibitem{Tolman}  R.C. Tolman, {\it Relativity, Thermodynamics and Cosmology},  
(Oxford, 1934).

\bibitem{Chandra} S. Chandrasekhar,
Mon. Not. R. Astron. Soc. {\bf 95}, (1935).

\bibitem{Einstein1911} A. Einstein,
Ann. der Phys. {\bf 35}, (1911), 898.

\bibitem{NFW96} J.F. Navarro, 
C.S. Frenk and D.M. White,  \AJ{462,1996,563}.

\bibitem{NFW97} J.F. Navarro, 
C.S. Frenk and D.M. White, \AJ{490,1997,493}.

\bibitem{M99} B. Moore, T. Quinn, F. Governato, 
J. Stadel and G. Lake, Mon. Not. R. Astron. Soc. {\bf 310}, (1999), 1147.  

\bibitem{F04} T. Fukushige, 
A. Kawai and J. Makino, 
\AJ{606,2004,625}.

\bibitem{Swaters03} R.A. Swaters, B.F. Madore,
F.C. van den Bosch and M. Balcells, \AJ{583,2003,732}.

\bibitem{deBlok02} W.J.G. de Blok and A. Bosma, 
Astron. Astrophys. {\bf 385}, (2002).

\bibitem{Gentile04} G. Gentile, P. Salucci, 
U. Klein, D. Vergani and P. Kaberla, Mon. Not. R. Astron. Soc. {\bf 351}, 
(2004) 903.

\bibitem{Gentile05} 
G. Gentile, A. Burkert, P. Salucci, U. Kelin and F. Walter,.
\AJ{634,2005,L145}.

\bibitem{JDSimon05} J.D. Simon, A.D. Bolatto, A. Leroy
and L. Blitz, \AJ{621,2005,757}.

\bibitem{Sofue01} Y. Sofue and V. Rubin, Ann. Rev.
Astron. Astrophys. {\bf 39}, (2001), 137.

\bibitem{Sofue99} Y. Sofue, Y. Tutui, M. Honma, 
A. Tomita, T. Takamiya, J. Koda and Y. Takeda, \AJ{523,1999,136}.

\bibitem{Kelson02} D.D. Kelson, A.I. Zabuldoff,
K.A. Williams, S.C. Trager, J.S. Mulchaey and M. Bolte, 
\AJ{576,720,2002}


\bibitem{Treu04} T. Treu and L.V.E. Koopmans,
\AJ{611,2004,739}.


\bibitem{Tyson98} J.A. Tyson,  
G.P. Kochanski and I.P. dell'Antonio, \AJ{498,1998,L107}.


\bibitem{TB05a}  T. Broadhurst {\it et al.}, \AJ{621,2005a,53}.

\bibitem{Smith01} G.P. Smith, J.-P. Kneib, H. Ebling,
O. Czoske and I.R. Smail, I. R., \AJ{552,2001,493}.

\bibitem{TB05b}  T. Broadhurst, M. Takada,
K. Umetsu, X.Kong, N. Arimoto, M. Chiba and T. Futamase,
\AJ{619,2005b,L143}.


\bibitem{Sand02} D.J. Sand, T. Treu and R.S. Ellis,
\AJ{574,2002,L129}.


\bibitem{Sand04} D.J. Sand, T. Treu, G.P. Smith
and R.S. Ellis, \AJ{604,2004,88}.




\end{thebibliography}
\end{document}